\documentclass[12pt,preprint]{aastex}

\shorttitle{Free-floating planets}
\shortauthors{Hurley \& Shara}

\slugcomment{To appear in Astrophysical Journal: submitted August, 1, 2001}

\begin{document}

\title{Free-Floating Planets: Not So Surprising}

\author{Jarrod R. Hurley and Michael M. Shara}
\affil{Department of Astrophysics, 
       American Museum of Natural History, 
       Central Park West at 79th Street, \\ 
       New York, NY 10024}
\email{jhurley@amnh.org, mshara@amnh.org}

\begin{abstract}
We find that free-floating planets can remain bound to a star cluster 
for much longer than was previously assumed: of the order of the 
cluster half-mass relaxation timescale as opposed to the crossing-time. 
This result is based on $N$-body simulations performed with the 
new GRAPE-6 special purpose hardware and is important 
in the context of the preliminary detection of a population of free-floating 
sub-stellar objects in the globular cluster M22. 
The planets in our $N$-body study are of Jupiter mass and are initially 
placed in circular orbits of between 0.05 and $50\,$AU about a parent 
star whose mass is chosen from a realistic initial mass function. 
The presence of the free-floating planets is the result of dynamical 
encounters between planetary systems and the cluster stars. 
Most planets are liberated from their parent star in, or near, the cluster 
core, and then drift outwards on a timescale of $\sim 10^{8-9}\,$year. 
This still implies the existence of many ($\sim 100$) planets per star 
if the M22 result is confirmed. 
\end{abstract}


\section{Introduction}
\label{s:intro}

The recent null detection of hot Jupiters in the globular cluster 
47 Tucanae \citep{gil00} is an important puzzle. 
This search for planetary systems was sensitive to the 
detection of gas-giant planets in orbits of less than five days 
($\sim 0.05\,$AU) about a main-sequence parent star: 
the so-called ``hot Jupiters''. 
If the frequency of hot Jupiters in the solar neighbourhood holds for 
47 Tuc then simulations suggested that approximately 20 should have been 
unearthed by the survey \citep{gil00}. 

The \citet{gil00} survey has led to a number of analytical and 
statistical studies concerning the fate of planetary systems in 
star cluster environments \citep{bon01,dav01,smi01}. 
\citet{dav01} estimate that only planets that form in orbits with 
semi-major axes, $a$, less than $0.3\,$AU will survive in a globular 
cluster but that even planets with $a \sim 0.04\,$AU would be broken-up 
in the high density core. 
In the case of 47 Tuc the question then seems to be: 
were all the surveyed stars resident in the cluster core for a 
significant fraction of the cluster evolution or, did the planets 
not form in the first place? 
\citet{bon01} have suggested that in the early globular cluster environment 
the natal disk from which planets form may be truncated inside the region 
where gas-giant planets are believed to form. 
In the solar neighbourhood the parent stars of planets tend to be 
considerably richer in metals than average \citep{lau00}.  
This lends support to the hypothesis that a lower abundance of metals in 
proto-planetary nebulae causes a lower frequency of planet formation as 
a result of fewer dust grains for nucleation 
\citep[and references within]{gil00}. 
The metallicity of 47 Tuc is a factor of five less than solar \citep{har96}. 

If we assume that a population of planets does form with orbital 
separations in the range $1 \leq a/{\rm AU} \leq 50$ then \citet{smi01} 
show that 50\% of these planets will be liberated from their parent star 
in a globular cluster and 27\% will be liberated in the less-dense 
surroundings of an open cluster. 
{\it However, these authors also claim that in an open cluster only 0.5\% 
of these liberated planets will stay in the cluster for more than 
a crossing-time.}  
This increases to 30.1\% for a typical globular cluster. 

In the case of M22 it is the liberated, or free-floating, planets that 
are of particular interest. 
The microlensing of background stars by compact objects in globular 
clusters has been analyzed in detail by \citet{pac94}. 
Possible targets mentioned for a Hubble Space Telescope (HST) study are 
M22 with the Galactic bulge background and 47 Tuc with the Small 
Magellanic Clouds as background. 
The advantage here is that the probability of lensing is high and 
accurate knowledge of the distances and kinematics of the sources and 
lenses leads to better lense-mass estimates. 
\citet{sah01} attribute six possible gravitational microlensing events 
to planetary-mass objects and only one event to a star, the mass of 
which is $0.13 M_{\odot}$, based on HST observations of M22.  
If these free-floating planetary-mass objects are Jupiters then their 
relative size suggests that there must be 60 Jupiters per star in M22 
and therefore that they constitute $\sim 10\%$ of the cluster mass 
\citep{pac01}. 
If they are Earth-like objects then the number increases to 600 per star,  
but they comprise only 0.3\% of the cluster mass.  
The central density of M22 is roughly $10^4\,$stars$\,{\rm pc}^{-3}$, a 
factor of 10 less than 47 Tuc, which still places it clearly within the 
globular cluster regime. 
Factoring in the 50\% liberation rate for planets effectively doubles 
the number of planets required per star. 
Furthermore, the relative time spent in the cluster by the liberated 
planets is then a critical factor in estimating the microlensing rate. 
Interestingly the metallicity of M22 is a factor of 10 lower than 
that of 47 Tuc \citep{har96}. 

A large fraction of the stars that we observe are found in gravitationally 
bound star clusters. 
In fact, it is entirely possible that all stars were born in a star 
cluster of some sort \citep{kra83,lad93}. 
Clusters are crowded stellar environments, ranging in density from 
$10^2\,$stars$\,{\rm pc}^{-3}$ to as high as $10^7\,$stars$\,{\rm pc}^{-3}$ 
in the cores of the densest globular clusters, which complicates matters 
for the evolution of their members. 
Encounters between stars can lead to collisions and, in the case of binary 
stars or planetary systems, an exchange interaction or disruption of the 
orbit. 
For this reason, amongst others, it is desirable to model the 
evolution of a star cluster using a 
direct $N$-body method in which the individual orbits of 
each star are followed in detail {\it and} the internal evolution of each star 
is also taken into account \citep{hur01}. 

We have instigated a study of the behaviour of planetary 
systems in star clusters using a state-of-the-art $N$-body code in 
conjunction with the powerful GRAPE-6 special purpose computer \citep{mak01}. 
This detailed project will ultimately involve a large number of 
$N$-body simulations covering a wide range of initial conditions, 
e.g. metallicity, binary fraction, stellar number density, 
and multiple planets per star. 
However, for now we have simply looked at the case of Jupiters in 
single-planet systems within moderate density cluster conditions. 
Furthermore, we do not consider whether planets should form at all in 
star clusters \citep[for example]{arm00}, especially in low-metallicity 
and/or high density environments, but simply ask the question: 
what happens if they do? 
Even though this project is in its infancy the possible discovery of 
free-floating planets in M22 \citep{sah01} makes publication of the 
initial results very timely.

\section{Simulation Method}
\label{s:method}

To model the evolution of star clusters we use the Aarseth {\tt NBODY4} 
code \citep{aar99,hur01}. 
Simulations are performed on a prototype GRAPE-6 board located at the 
American Museum of Natural History. 
This special purpose hardware, which acts as a Newtonian force accelerator 
for $N$-body calculations, performs $0.5\,$Tflops 
($\sim 30\,$Gflop per chip). 
It represents a factor of 100 increase in computing power compared to 
its predecessor the GRAPE-4 \citep{mak93} and has brought the possibility of 
modelling globular clusters on a star-to-star basis within reach for 
the first time\footnote{We refer the interested reader to 
astrogrape.org for further information on the GRAPE project.}.  

The simulations performed so far have involved $22\,000$ stars with 
a 10\% primordial binary fraction. 
Initial conditions relating to the masses, positions and velocities 
of the stars, as well as the orbital characteristics of the binaries, 
are the same as for the $N = 10\,000$ star simulations described in 
detail by \citet{hur01}. 
In particular, a realistic initial-mass function is used to distribute 
the stellar masses \citep{kro93}, and the cluster is subject to a 
standard Galactic tidal field. 
The distribution of orbital separations for the primordial binaries is 
log-normal with a peak at $30\,$AU, and spans the range 
$\sim 6 \, R_{\odot}$ to $30\,000\,$AU. 
The eccentricity of each binary orbit is taken from a thermal distribution 
\citep{heg75}. 
Positions and velocities of the stars are assigned according to a 
Plummer model \citep{aar74} in virial equilibrium. 

We include the outcome of three simulations in the results presented here 
(see Table~\ref{t:table2}). 
The first had a metallicity of $Z = 0.004$, relevant to 47 Tuc, and 
included $2\,000$ planets of Jupiter mass. 
Each planet was placed in a circular orbit about a randomly chosen parent 
star at a separation taken from a uniform distribution between 1 and  
$50\,$AU.  
The second simulation involved $3\,000$ Jupiters with the lower limit of 
the separation distribution reduced to $0.05\,$AU and the final simulation 
differs from this only in the use of $Z = 0.02$. 
Each simulation was evolved to an age of $4.5\,$Gyr when $\sim 25\%$ of the 
initial cluster mass remained and the binary fraction was still close 
to 10\%. 
Typically the velocity dispersion of the stars in these model clusters 
was $2\,{\rm km } \,{\rm s}^{-1}$ with a core density of 
$10^3\,$stars$\,{\rm pc}^{-3}$. 
The density of stars at the half-mass radius is generally a factor of 
10 less than this. 

\section{Free-Floating Planets}
\label{s:result}

Table~\ref{t:table3} shows, as a function of time, the number of planets 
that are liberated from their parent stars during the simulation, 
the number of planetary systems that escape from the cluster, 
and the number of planets that are 
exchanged from their original orbit into orbits about another parent star. 
Also shown are the number of planets swallowed by their parent star 
simply as a result of nuclear driven expansion of the stellar envelope. 
These are averaged results from the three simulations. 

\begin{figure}
\plotone{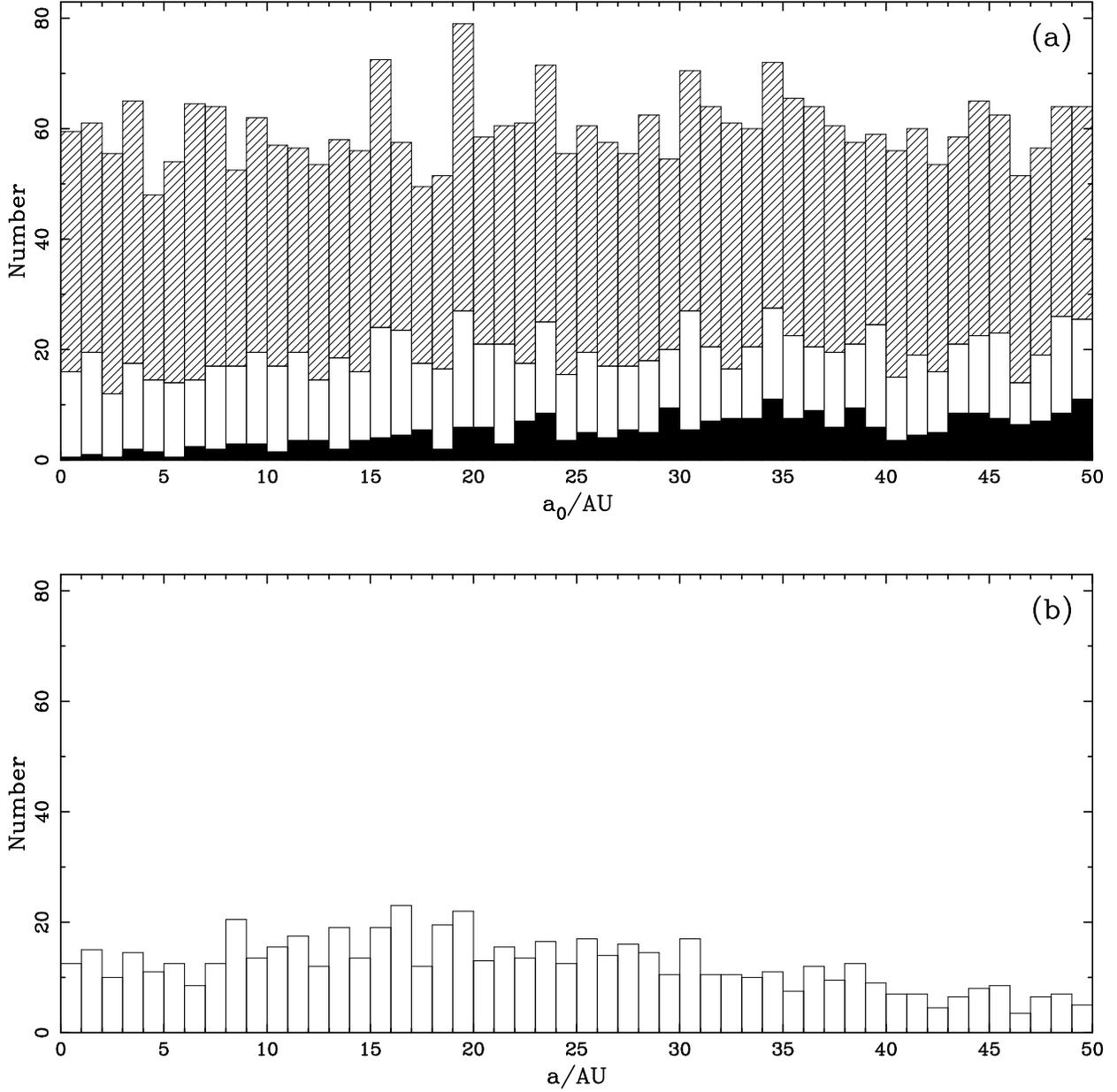}
\caption{Histograms relating to the planetary system separations: 
(a) the total of the distribution represents the initial planetary systems, 
    the solid region represents those that have been broken-up, and the 
    hatched region shows the number of planetary systems that have escaped; 
(b) the distribution of separations for planetary systems that remain in 
    the cluster when the simulation ended. 
Differences between the unshaded regions in (a) and (b) are attributed to 
exchange interactions, orbital changes owing to mass-loss or weak 
perturbations, and mergers of planets with their parent star. 
Note that only results from the second and third simulations are presented 
in this figure. 
Numbers in each bin are averaged over these two simulations. 
\label{f:fig1}}
\end{figure}

We find a weak preference for planets in wide orbits to be liberated 
from their parent star: planetary systems with a $50\,$AU separation 
are 10 times more likely to be broken-up than those with $1\,$AU 
(see Figure~\ref{f:fig1}). 
\citet{heg96} showed that the cross-section for a binary to undergo an 
exchange interaction, which also serves as a likelihood-of-disruption 
indicator, scales linearly with the orbital separation. 
This has been confirmed by \citet{dav01} in the case of planetary systems. 
The fact that we do not observe this relation is primarily a result of 
the large fraction of escaping systems which deprives the cluster of 
orbits to break-up. 
Another factor is the relatively weak binding energy of the planetary systems 
compared to that of the binaries. 
It is evident from Figure~\ref{f:fig1} that we are limited at this stage 
to a fairly low number of systems per orbital separation bin and not 
until we can saturate the distribution with a large number of systems, all 
situated in the core of the cluster, will we be able to fully test the 
statistical results mentioned above. 

Planetary systems primarily escape from a cluster owing to stripping 
of stars in the outer cluster regions by the Galactic tidal field. 
As a natural consequence of mass-segregation there is a preference for 
systems with low parent star mass to escape. 
Planetary systems of all orbital separations are equally likely to escape 
(as planets just ``tag along for the ride'', see Figure~\ref{f:fig1}). 
It is also possible for stars to be ejected from the cluster due to close 
encounters with other stars or binaries but in the case of planetary 
systems the encounter more likely results in liberation of the planet. 

\begin{figure}
\plotone{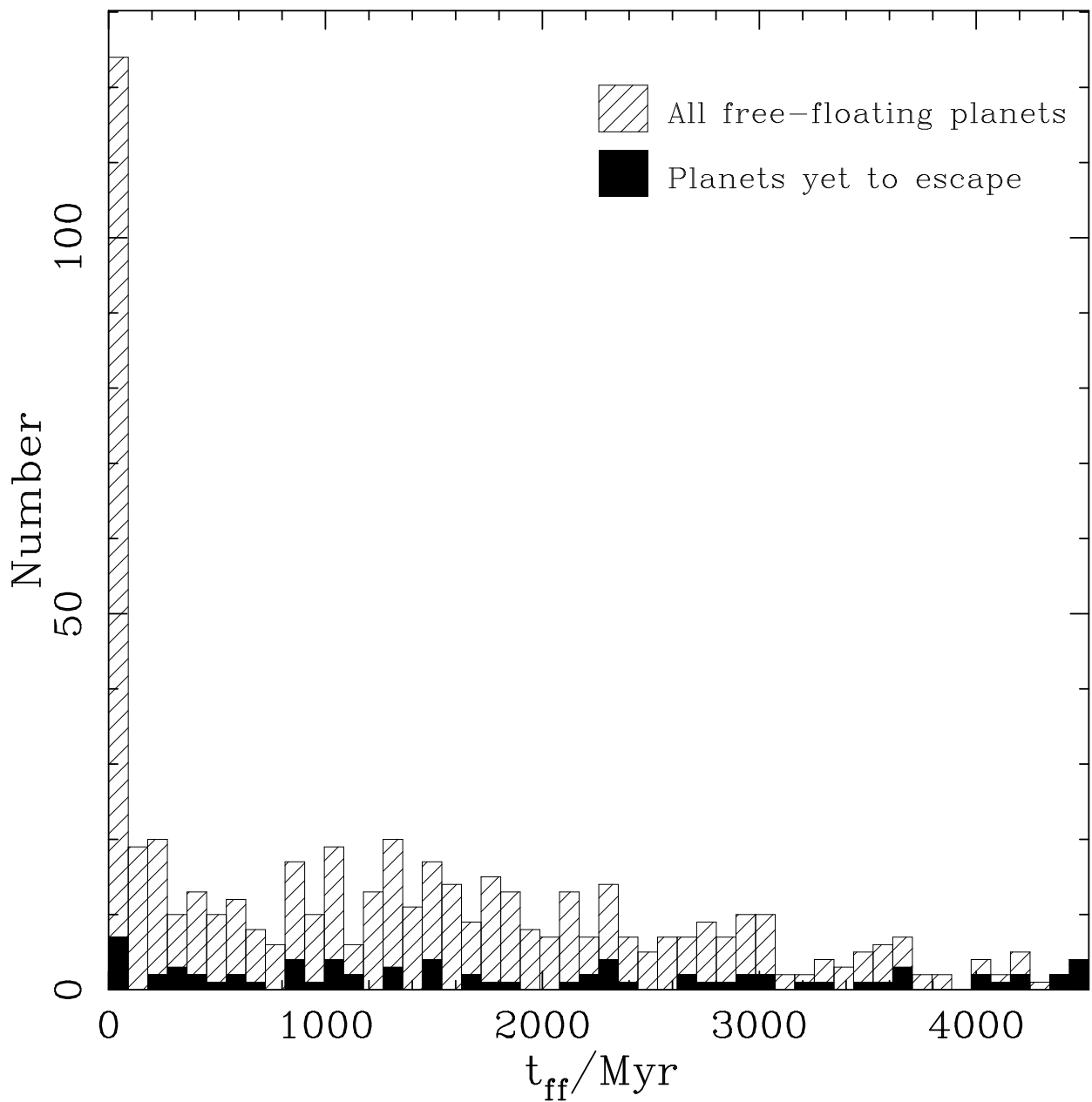}
\caption{Histogram showing the distribution of the time spent in the 
cluster by free-floating planets subsequent to liberation from their 
parent star (hatched region). 
The distribution for the subset of planets that remain in 
the cluster when the simulation was halted is also shown (solid region). 
For these planets 
the data represents a lower limit to the time spent in the cluster. 
\label{f:fig2}}
\end{figure}

\begin{figure}
\plotone{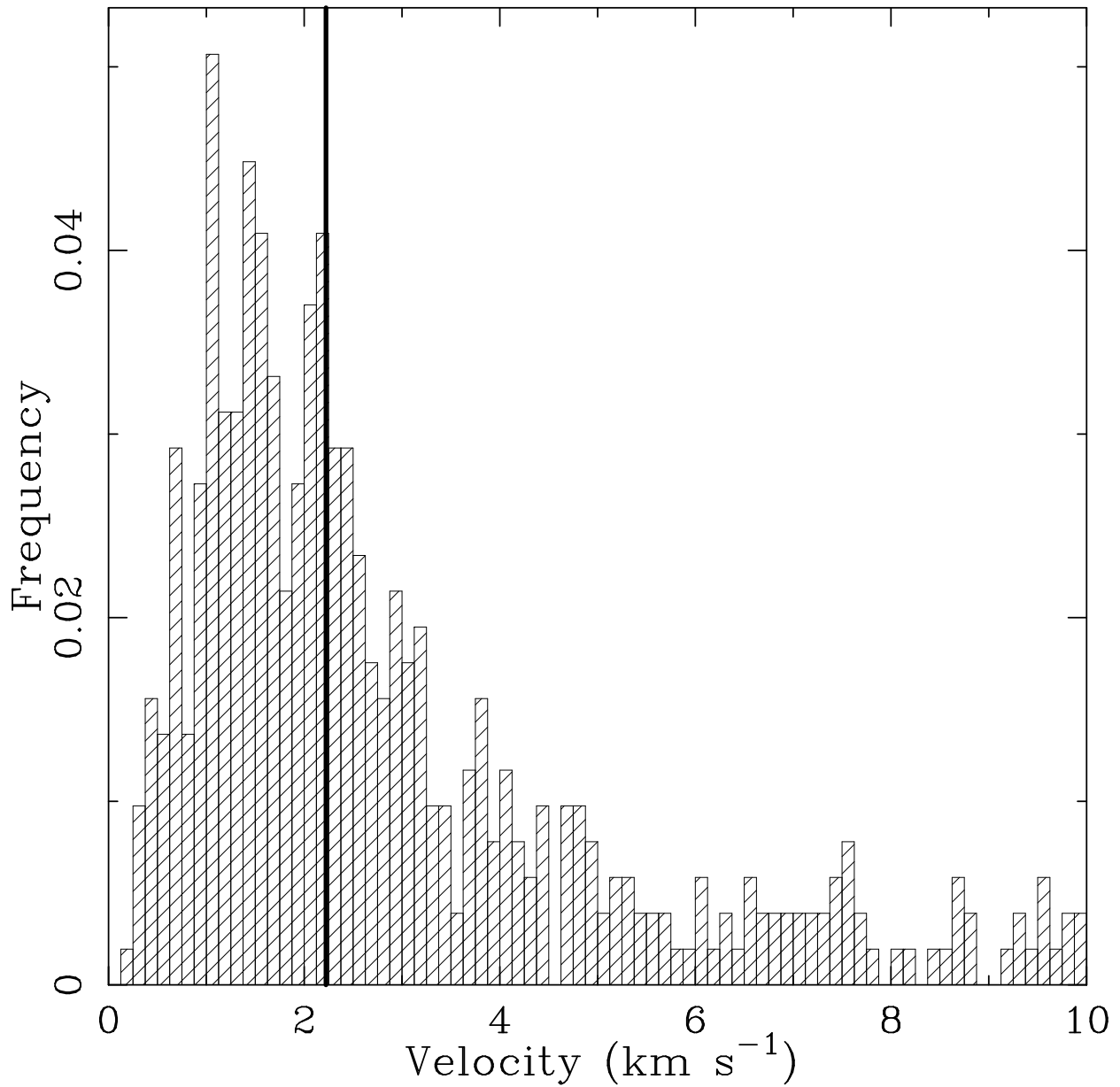}
\caption{Distribution of velocities for free-floating planets immediately 
after being liberated from their parent star. 
The distribution is normalized to the total number of liberated planets. 
The average cluster escape velocity is also shown (solid vertical line): 
46\% of planets are liberated at speeds lower than the cluster escape 
velocity.
The tail of the distribution is truncated at $10\,{\rm km } \,{\rm s}^{-1}$ 
which excludes the 10\% of the liberated planets with velocities 
extending out to $70\,{\rm km } \,{\rm s}^{-1}$. 
\label{f:fig3}}
\end{figure}

{\it We find that a large fraction of the liberated planets are retained in 
the cluster for much longer than a crossing-time.}  
The typical crossing-time for these simulations is $2-10\,$Myr. 
Figure~\ref{f:fig2} shows the distribution of time spent in the cluster 
by the free-floating planets. 
The planets are preferentially liberated in the cluster core and 46\% 
are liberated with a velocity less than the cluster escape velocity 
(see Figure~\ref{f:fig3}). 
The velocity dispersion of the free-floating planets is approximately 
twice that of the cluster stars. 
We expect this to have only a minimal effect on the determination of the 
lensing mass in M22 \citep{sah01}. 
So the planets generally begin their free-floating existence deep within 
the potential well of the cluster and will then journey towards 
the outer regions of the cluster driven by the effect of two-body 
relaxation. 
\citet{che90} derive the timescale for mass-segregation to be directly 
related to the relaxation timescale of the cluster but with an inverse 
dependence on stellar mass. 
Therefore we would expect the planets to take much longer to reach the tidal 
boundary of the cluster than low-mass stars. 

\begin{figure}
\plotone{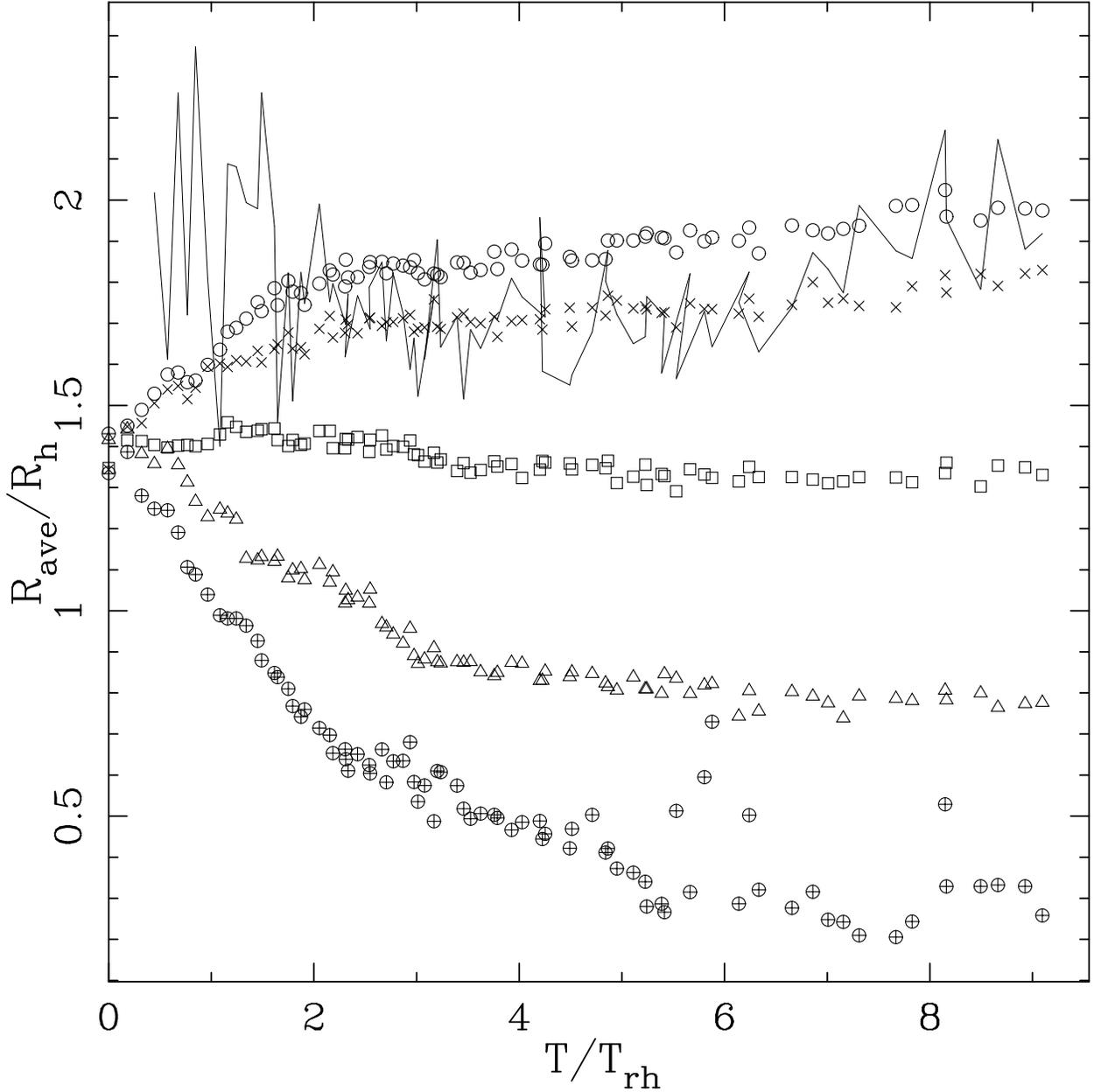}
\caption{The average radial position, scaled by the cluster half-mass 
radius, of various mass groups as a function of time, scaled by the 
current half-mass relaxation timescale, $T_{\rm rh}$. 
The mass groups identified are: $0.1-0.2 M_{\odot}$ ($\circ$), 
$0.2-0.5 M_{\odot}$ ($\times$), $0.5-1.0 M_{\odot}$ ($\Box$), 
$1.0-1.6 M_{\odot}$ ($\triangle$), $1.6-3.0 M_{\odot}$ ($\oplus$), 
and the free-floating planets (solid line). 
The planet group contains significantly fewer members, approximately 50 at 
any one time, which explains the increased noise in its data, while data  
for the most massive group becomes noisy after about $300\,$Myr when its 
members start to evolve off the main-sequence.  
The half-mass relaxation timescale is typically within the range of 
$200-400\,$Myr during the cluster lifetime. 
\label{f:fig4}}
\end{figure}

This is not what we see in Figure~\ref{f:fig4} which illustrates the average 
position within the cluster over time for various mass groups. 
As expected the $0.5-1.0 M_{\odot}$ group, which always contains the average 
stellar mass, shows little movement. 
For the remaining stellar mass groups there is a strong correlation 
between deviation from the average stellar mass and the rate of 
mass-segregation, whether it be inwards for high-mass or outwards 
for low-mass.
This clearly demonstrates that equipartition of energy is dominating 
the dynamical evolution. 
The picture is complicated for the planets because, in this case, the core 
population is replenished over time and their velocity distribution is 
detached from that of the stars. 
From Figure~\ref{f:fig4} we see that the average position of the free-floating 
planet population remains roughly constant, lying just outside the 
half-mass radius. 
The planets take approximately $200\,$Myr, comparable to the half-mass 
relaxation timescale of the cluster, to move from inside the core to 
outside the half-mass radius. 

For the non-escaping planetary systems we find that marginally more planets 
are liberated than suggested by \citet[$\sim 30\%$ compared to 27\%]{smi01} 
and that a much larger fraction of free-floating planets are retained 
($\sim 64\%$ compared to 0.5\%). 
If this trend continues into the globular cluster regime, and we expect 
that it will, then these early results have an important bearing on the 
interpretation of the M22 observations.   
In particular it is not as surprising as one might have originally 
thought that there are many free-floating planets in M22. 

We note that \citet{smi01} only considered equal-mass stars: 
they studied cases of $0.7 M_{\odot}$ and $1.5 M_{\odot}$ separately.  
In this respect the velocity distribution shown in Figure~\ref{f:fig3} 
is more representative of the real picture and this goes some way 
towards explaining our vastly different results for the retention of 
planets. 
\citet{smi01} also note that changes in the planet mass can 
critically affect the velocity distribution. 
Table~\ref{t:table3} shows that the current number of planets in 
the cluster is decreasing with time but the escaping planetary 
systems are distorting the results. 
For the same reason our numbers concerning the liberation of planets 
must be taken as lower limits. 
However, when comparing these numbers to the work of \citet{smi01} 
we note that they used a separation distribution uniform in log space 
for their planetary systems and thus we have a higher proportion of 
initially wide orbits. 
The effect of escaping systems must be addressed in future simulations, 
possibly by placing planets primarily around stars with mass close to the  
average for the cluster stars. 

\section{Planetary Orbits}
\label{s:orbits}

A significant number of planetary orbits are altered during the simulations. 
Orbits of all sizes expand when the parent star evolves off the 
main-sequence and begins to lose mass non-conservatively in a stellar 
wind. 
Weak perturbations from passing stars can cause the orbital period to 
decrease: this affects roughly 15\% of systems with $a > 10\,$AU. 
We do not see any orbital migration of planetary systems with initial 
orbital separations less than this. 
A {\it hard} system is defined as having an orbit with a binding energy 
greater than that of the mean kinetic energy of the cluster stars 
\citep{heg75}. 
It is then expected that owing to close encounters during the cluster 
evolution hard systems will become harder, 
i.e. orbital migration inwards, and {\it soft} systems will be broken-up. 
The hard/soft limit for binaries in our cluster simulations is 
roughly $60\,$AU and for the planetary systems it is more like $0.1\,$AU. 
Considering this in conjunction with the relatively low number density 
of stars in the simulations performed so far it is not surprising that 
we have yet to observe hardening of close planetary orbits. 
Exchange interactions alter the observed distribution of orbital 
characteristics in a fairly random manner although it is more likely 
for a wide system to be involved in such an event.  

\section{Conclusions}
\label{s:conclu}

Contrary to recent claims \citep{bon01,smi01} we find that free-floating 
planets can form a significant population in stellar clusters. 
This is based on the results of open cluster size $N$-body simulations 
but is expected to be even more likely in the case of globular clusters. 
While it should be stressed that the detection of free-floating planets 
in M22 is preliminary, and also speculative, 
it suggests that at least 100 planets were formed for every star. 
This may sound implausible but is in fact supported by recent simulations.  
\citet{ida01} have shown that in a protoplanetary disk where the surface 
density of the solid component is low, the isolation mass of planets 
is small and many terrestial planets can form. 
It is also possible that protoplanetary disks having lower metallicity 
than solar would form many earth-like planets - perhaps  
50-100 per star (Shigeru Ida, private communication).
A population of free-floating sub-stellar objects has also been detected 
in the young cluster $\sigma$ Orionis \citep{zap00}. 
The possibility has been raised that these may be formed as such 
\citep{bos01}, i.e. not attached to a parent star. 

We agree with \citet{dav01} that subsequent surveys for planetary systems 
should be conducted in clusters less dense that 47 Tuc, such as metal-rich 
open clusters. 
Observations of a metal-rich globular cluster should 
help determine whether the lack of planetary systems in 47 Tuc is due 
to the metallicity of the cluster or dynamical interactions. 
We note that a planet has been detected within a binary pulsar system 
in M4 \citep{tho99} which is metal-poor compared to 47 Tuc \citep{har96}. 

As we expand the parameter space of our $N$-body study many of the 
interesting issues regarding planetary systems in star clusters will 
be addressed. 
Of particular importance will be the inclusion of systems with multiple 
planets per star \citep{mur01}. 
Full realisation of the capabilities of the GRAPE-6 hardware when the 
$1\,$Tflop board becomes available will allow larger particle numbers, 
and consequently more planetary systems, to be studied per simulation. 
This will improve the statistical significance of our results considerably. 
Moving to simulations that operate at globular cluster number densities 
will make it possible to look for orbital migration in small-period 
planetary systems. 
Hence this study will have implications for future planet searches in 
globular clusters, especially if hot Jupiter planetary systems cannot 
form directly in such an environment.

\acknowledgments

We are extremely grateful to Jun Makino and the University of Tokyo 
for the loan of the GRAPE-6 board. 
We thank David Zurek, Shigeru Ida and Rosemary Mardling 
for helpful discussions.

\clearpage

\clearpage

\begin{table}
\begin{center}
\caption{ 
Parameters of the $N$-body simulations presented in this work. 
The simulation ID number, metallicity, number of planetary systems, 
and the minimum and maximum of the planetary orbital separation distribution 
are listed. 
Each simulation involved $22\,000$ stars, comprised of $18\,000$ single 
stars and $2\,000$ binaries, so the total number of particles in each 
simulation is $22\,000 + N_{\rm p}$. 
\label{t:table2}}
\begin{tabular}{ccccc}
\tableline
Run & $Z$ & $N_{\rm p}$ & $a_{\rm min}$ & $a_{\rm max}$ \\ 
\tableline
1 & 0.004 & 2000 & 1.00 & 50.0 \\
2 & 0.004 & 3000 & 0.05 & 50.0 \\
3 & 0.020 & 3000 & 0.05 & 50.0 \\
\tableline
\end{tabular}
\end{center}
\end{table}

\begin{table}
\begin{center}
\caption{Averaged results for the planet population at $1.0\,$Gyr 
intervals. The percentage of all planets liberated from their parent star, 
the percentage of these that remain in the cluster for more than a 
crossing-time, and the percentage in the cluster at that time, are given 
in columns 2-4. The percentage that have escaped attached to their parent star 
is given in column 5. Columns 6 and 7 give the percentages of planets that 
have been engulfed by their parent star, and those that have been exchanged 
into orbit about another parent star, respectively. 
\label{t:table3}}
\begin{tabular}{c|rrr|r|r|r}
\tableline\tableline
Time/Myr & & Liberated & & Escaped & Swallowed & Exchanged \\ 
 & Total & Kept & Current & & & \\ 
\tableline
1000.0 & 5.6\% & 69.8\% & 48.4\% & 11.7\% & 0.4\% & 1.0\% \\
2000.0 & 7.7\% & 66.8\% & 33.1\% & 31.4\% & 0.8\% & 1.7\% \\
3000.0 & 9.1\% & 66.5\% & 22.4\% & 51.1\% & 0.9\% & 2.3\% \\
4000.0 & 10.4\% & 64.0\% & 12.7\% & 65.8\% & 1.0\% & 3.6\% \\ 
\tableline\tableline
\end{tabular}
\end{center}
\end{table}


\newpage 

\begin{thebibliography}{}
\bibitem[Aarseth, H\'{e}non \& Wielen (1974)]{aar74} Aarseth, S.,
    H\'{e}non, M., \& Wielen, R. 1974, \aa, 37, 183 
\bibitem[Aarseth (1999)]{aar99} Aarseth, S. J. 1999, \pasp, 111, 1333 
\bibitem[Armitage (2000)]{arm00} Armitage, P. J. 2000, \aap, 362, 968 
\bibitem[Bonnell et al. (2001)]{bon01} Bonnell, I. A., Smith, K. W., 
    Davies, M. B., \& Horne, K. 2001, \mnras, 322, 859 
\bibitem[Boss (2001)]{bos01} Boss, A. P. 2001, \apj, 551, L167
\bibitem[Chernoff \& Weinberg (1990)]{che90} Chernoff, D. F., 
    \& Weinberg, M. D. 1990, \apj, 351, 121 
\bibitem[Davies \& Sigurdsson (2001)]{dav01} Davies, M. B., 
    \& Sigurdsson, S. 2001, \mnras, 324, 612 
\bibitem[Gilliland at al. (2000)]{gil00} Gilliland, R. L., et al. 2000, 
    \apj, 545, L47 
\bibitem[Harris (1996)]{har96} Harris, W. E. 1996, \aj, 112, 1487
\bibitem[Heggie (1975)]{heg75} Heggie, D.C. 1975, \mnras, 173, 729 
\bibitem[Heggie, Hut \& McMillan (1996)]{heg96} Heggie, D. C., Hut, P., 
    \& McMillan, S. L. W. 1996, \apj, 467, 359  
\bibitem[Hurley et al. (2001)]{hur01} Hurley, J. R., Tout, C. A., 
    Aarseth, S. J., \& Pols, O.R. 2001, \mnras, 323, 630
\bibitem[Ida \& Kokubo (2001)]{ida01} Ida, S., \& Kokubo, E. 2001, 
    in ASP Conf. Ser. XX, Planetary System in the Universe: 
    Observation, Formation \& Evolution, ed.  A. J. Penny, P. Artymowicz, 
    A. -M. Lagrange, \& S. S. Russell (San Francisco: ASP), in press 
\bibitem[Kraft (1983)]{kra83} Kraft, R. P. 1983, 
    Highlights in Astronomy, 6, 129 
\bibitem[Kroupa, Tout \& Gilmore (1993)]{kro93} Kroupa, P., Tout, C. A., 
    \& Gilmore, G. 1993, \mnras, 262, 545 
\bibitem[Lada, Strom \& Myers (1993)]{lad93} Lada, E. A., Strom, K. M., 
    \& Myers, P. C. 1993, in Protostars and Planets III, ed. 
    E. Levy, \& J. Lunine (University of Arizona), 245 
\bibitem[Laughlin (2000)]{lau00} Laughlin, G. 2000, \apj, 545, 1064 
\bibitem[Makino, Kokubo \& Taiji (1993)]{mak93} Makino, J., Kokubo, E., 
    \& Taiji, M. 1993, \pasj, 45, 349 
\bibitem[Makino (2001)]{mak01} Makino, J. 2001, in ASP Conf. Ser. XX, 
    Stellar Collisions, ed. M. M. Shara (San Francisco: ASP), in press 
\bibitem[Murray \& Holman (2001)]{mur01} Murray, N., \& Holman, M. 2001, 
    \nat, 410, 773 
\bibitem[Paczynski (1994)]{pac94} Paczynski, B. 1994, Acta Astronomica, 44, 235 
\bibitem[Paczynski (2001)]{pac01} Paczynski, B. 2001, \nat, 411, 1002 
\bibitem[Sahu et al. (2001)]{sah01} Sahu, K. C., Casertano, S., Livio, M., 
    Gilliland, R. L., Panagia, N., Albrow, M. D., \& Potter, M. 2001, 
    \nat, 411, 1022 
\bibitem[Smith \& Bonnell (2001)]{smi01} Smith, K. W., \& Bonnell, I. A. 
    2001, \mnras, 322, L1
\bibitem[Thorsett et al. (1999)]{tho99} Thorsett, S. E., Arzoumanian, Z., 
    Camilo, F., \& Lyne, A. G. 1999, \apj, 532, 763 
\bibitem[Zapatero-Osorio et al. (2000)]{zap00} Zapatero-Osorio, M. R., 
    B\'{e}jar, V. J. S., Mart\'{i}n, E. L., Rebolo, R., 
    Barado y Navascue\'{e}s, D., Bailer-Jones, C. A. L., \& Mundt, R. 2000, 
    Science, 290, 103
\end{thebibliography}
\end{document}